\documentclass[a4paper,11pt]{article}
\usepackage{pos}

\usepackage{graphicx}
\usepackage[normalem]{ulem}

\title{Evidence for and implications of a dark photon}
\author*[a]{X.~G.~Wang}
\author[a]{A.~W.~Thomas}

\affiliation[a]{CSSM and ARC Centre of Excellence for Dark Matter Particle Physics,\\ 
Department of Physics, University of Adelaide, Adelaide, SA 5005, Australia}

\emailAdd{xuan-gong.wang@adelaide.edu.au}
\emailAdd{anthony.thomas@adelaide.edu.au}

\abstract{
We performed the first global QCD analysis of electron-nucleon deep-inelastic scattering and related high-energy data by including the contribution from a dark photon. Our results revealed a significant reduction in $\chi^2$ relative to the baseline result without new physics. From a hypothesis test, our best dark photon fit is preferred over the Standard Model by as much as $6.5\ \sigma$, providing the first hint for the existence of a dark photon, although indirectly. Additionally, we explored the implications of a dark photon in party-violating electron scattering, rare kaon decay, and the electroweak precision observables. The dark photon as a portal connecting to dark matter particles was also discussed.
}

\FullConference{The 21st International Conference on Hadron Spectroscopy and Structure (HADRON2025)\\
 27 - 31 March, 2025\\
Osaka University, Japan\\}

\begin{document}
\maketitle

\section{Introduction}
Despite great success, the Standard Model (SM) is still incomplete as it involves a large number of free parameters, does not explain the neutrino masses or oscillations and does not incorporate dark matter. In addition, there are many potential new physics phenomena, such as anomalies in the muon $g-2$~\cite{Muong-2:2023cdq}, the $W$ boson mass~\cite{CDF:2022hxs}, and the rare kaon~\cite{NA62:2024pjp} and $B$ meson decays~\cite{Belle-II:2023esi}. Although still needing further confirmation~\cite{Boccaletti:2024guq, CMS:2024lrd}, these have inspired new physics searches both theoretically and experimentally.

There are many popular extensions to the Standard Model in the gauge sector (extra $Z'$ or $W'$), the Higgs sector (two-Higgs-Doublet model), and the neutrino sector (sterile neutrinos). Among these well-motivated new physics models, the extensions with an extra $U(1)$ gauge field have received considerable attention. It is usually introduced through anomaly free $U(1)'$ charges in the $Z'$ model~\cite{Appelquist:2002mw, Leike:1998wr},
\begin{equation}
D_{\mu} \psi_L = (\partial_{\mu} + i g \boldsymbol{A}_{\mu}\cdot \boldsymbol{t} + \frac{1}{2} i g' Y B_{\mu} +  \frac{1}{2} i g_z z_{\psi} B'_{\mu}) \psi_L\, ,
\end{equation}
where the anomaly cancellation conditions impose strong constraints on $z_{\psi}$, or through kinetic mixing with the SM hypercharge~\cite{Fayet:1980ad, Fayet:1980rr, Holdom:1985ag} in the dark photon model,
\begin{eqnarray}
\label{eq:L_DP}
{\cal L}_{\rm DP} & = & 
- \frac{1}{4} F'_{\mu\nu} F'^{\mu\nu} + \frac{1}{2} m^2_{A'} A'_{\mu} A'^{\mu} 
+ \frac{\epsilon}{2 \cos\theta_W} F'_{\mu\nu} B^{\mu\nu} 
\, ,
\end{eqnarray}
where $\epsilon$ is the mixing parameter.

 We will focus on the dark photon model. 
 With the inclusion of a dark photon, the $Z$ boson mass and its weak couplings will be shifted with respect to their SM predictions. Moreover, the dark photon itself will interact with the SM fermions with both vector and axial-vector couplings.
 
While numerous experimental searches have been undertaken, no direct signal of a dark photon has been found so far~\cite{BaBar:2014zli, LHCb:2019vmc, CMS:2019buh}. Rather stringent exclusion constraints have been placed on the mixing parameter, leading to upper limits of $\epsilon \le {\cal O}(10^{-3})$ over a wide range of dark photon masses~\cite{BaBar:2014zli, CMS:2019buh}. Constraints on $\epsilon$ from theoretical investigations in connection with the muon $g-2$~\cite{Pospelov:2008zw}, electroweak precision observables~\cite{Hook:2010tw, Curtin:2014cca}, and electron-nucleon deep inelastic scattering (DIS)~\cite{Kribs:2020vyk, Thomas:2021lub, Yan:2022npz} are much weaker, with upper bounds being of ${\cal O}(10^{-2}) \sim {\cal O}(10^{-1})$.

 In Section~\ref{sec:evidence} we describe the first hint, albeit indirect, for the existence of a dark photon based upon a global QCD analysis~\cite{Hunt-Smith:2023sdz} of deep-inelastic scattering data. We explore the implications of a dark photon in Section~\ref{sec:implications}, including party-violating electron scattering~\cite{Thomas:2022qhj}, rare kaon decay~\cite{Wang:2023css}, and electroweak precision observables~\cite{Loizos:2023xbj}. A summary of our conclusions is given in Section~\ref{sec:conclusion}.

%%%%%%%%%%%%%%%%%%%%%%%%%%%%%%%%%%%%%%%%%%%%%%%%%%%%%%
\section{First hint of the existence of a dark photon}
\label{sec:evidence}
The electron-nucleon deep inelastic scattering (DIS) and related high-energy scattering data are attractive in probing new physics because of the large dataset and wide kinematic coverage.
With the inclusion of a dark photon, the $F_2$ and $F_3$ structure functions of the proton are given by~\cite{Kribs:2020vyk, Hunt-Smith:2023sdz}
\begin{equation}
    \begin{aligned}
        & \widetilde{F}{}_{2} = \sum_{i,j = \gamma,Z,A_D} \kappa_i \kappa_j F^{ij}_2, \\
        & \widetilde{F}{}_{3} = \sum_{i,j = \gamma,Z,A_D} \kappa_i \kappa_j F^{ij}_3,
    \end{aligned}
\end{equation}
where $\kappa_i = Q^2/(Q^2 + m_i^2)$.
At leading order (LO) in the strong coupling $\alpha_s$, one has
\begin{equation}
    \begin{aligned}
        & F^{ij}_2 = \sum_q (C^v_{i,e} C^v_{j,e} + C^a_{i,e} C^a_{j,e})(C^v_{i,q} C^v_{j,q} + C^a_{i,q} C^a_{j,q})\, x f_q, \\
        & F^{ij}_3 = \sum_q (C^v_{i,e} C^a_{j,e} + C^a_{i,e} C^v_{j,e})(C^v_{i,q} C^a_{j,q} + C^a_{i,q} C^v_{j,q})\, x f_q,
    \end{aligned}
\end{equation}
where $C^{v(a)}_{i,f}$ is the vector (axial-vector) coupling of the physical gauge boson $i$ to the SM fermion $f$, $x$ is the parton momentum fraction, and $f_q$ is the parton distribution function (PDF) for quark flavor $q$ in the proton. At the initial scale $Q^2_0$, the PDFs can be parametrised as
\begin{equation}
  f_q(x,Q^2_0) = N x^{\alpha} (1 - x)^{\beta} ( 1 + \gamma \sqrt{x} + \eta x)\ , 
\end{equation}
where the free parameters are determined from a global fit analysis.

In our work~\cite{Hunt-Smith:2023sdz}, we used the JAM (Jefferson Lab Angular Momentum) framework at next-to-leading order (NLO) in $\alpha_s$, which employs Monte Carlo techniques and state-of-the-art uncertainty quantification~\cite{Cocuzza:2021cbi}. We also took into account potential missing higher order uncertainties,  following the procedure developed by the NNPDF collaboration~\cite{NNPDF:2019ubu}.

We first performed a baseline fit without new physics. The values of $\chi^2$ per degree of freedom for various datastes are given in the third column of Tab.~\ref{table:chi2}. We then included the dark photon contributions and repeated the global fit analysis. We found a significant reduction in the total $\chi^2$ with respect to the baseline result~\cite{Hunt-Smith:2023sdz}, mainly coming from the fixed target DIS and HERA neutral current (NC) processes which have large numbers of data points. The best dark photon fit gave $M_{A_D} \in (4, 6)\ {\rm GeV}$ and $\epsilon \in (0.06,0.12)$ at 95\% CL.
However, the improvement in $\chi^2$ is so substantial that if we perform the hypothesis test with $M_{A_D}=3\ {\rm GeV}$ and $\epsilon = 0.03$, the dark photon model is still preferred over the SM with a significance above $4\ \sigma$. Our results provided the first hint for the existence of a dark photon, albeit indirectly.

\begin{table}[!h]
\begin{center}
% \begin{tabular}{ m{8em} |c|c|r } 
\begin{tabular}{l|c|c|r}\hline\hline
    ~reaction & ~$\chi^2_{\rm dof}({\rm dark})$~ & ~$\chi^2_{\rm dof}({\rm baseline)}$~ & ~$N_{\rm dof}$~ \\
    \hline 
    ~fixed target DIS & 1.01 & 1.05 & 1495~~\\
    ~HERA NC          & 1.02 & 1.03 & 1104~~\\
    ~HERA CC          & 1.13 & 1.18 & 81~ \\
    ~Drell-Yan        & 1.18 & 1.16 & 205~ \\ 
    ~$Z$ rapidity     & 1.08 & 1.05 & 56~ \\ 
    ~$W$ asymmetry    & 1.04 & 1.07 & 97~ \\ 
    ~jets             & 1.16 & 1.15 & 200~ \\ 
    \hline
    ~\bf{total} & 1.03 & 1.05 & 3283~ \\ \hline
\end{tabular}
\caption{The $\chi^2$ values per degree of freedom with (``dark'') and without (``baseline'') dark photon modifications for various datasets, taking into account potential missing higher order uncertainties.}
\label{table:chi2}
\end{center}
\end{table}

In the light of this result it was important to test  whether the improvement in $\chi^2$ still holds for other new gauge bosons. This led us to include the $U(1)_{B-L}$ $Z'$ boson in a global QCD analysis~\cite{Wang:2024gvt}. For the electron-nucleon DIS process, the main differences between the $Z'$ model and the dark photon model come from their couplings to the SM fermions.
Contrary to the dark photon case, the minimum $\chi^2$ found with a $Z'$ boson was always worse with respect to the baseline fit, meaning that the $U(1)_{B-L}$ $Z'$ model was not favoured by the high-energy scattering data. This result clearly showed that the inclusion of new gauge bosons does not guarantee an improvement in the $\chi^2$ in a global QCD analysis.

Another question is that the mixing parameter from our best fit is in strong tension with the upper limits set by direct experimental searches. Regarding this, we pointed out that the dark photon signal in direct searches is model dependent, and the exclusion constraints on the mixing parameter could be significantly relaxed in light of potential couplings of the dark photon to dark matter particles~\cite{Felix:2025afw}. 
Taking the CMS constraints~\cite{CMS:2019buh} as an example, the dark photon contributions to $q\bar{q} \to \mu^+ \mu^-$ depend on its total decay width within a Breit-Wigner parametrisation,
\begin{equation}
 \sigma_{A_D} \propto \frac{\Gamma_{A_D \to q \bar{q}} \cdot \Gamma_{A_D \to \mu^+\mu^-}}{(s - M^2_{A_D})^2 + s \Gamma^2_{A_D}(s)}\, ,  
\end{equation}
where $\Gamma_{A_D}(s) = \sqrt{s} \Gamma_{A_D}/M_{A_D}$.
A narrow resonant peak is expected in the di-muon invariant mass spectrum by assuming that the dark photon only decays to the SM final states in which case its total width is suppressed by $\epsilon^2$. However, if the dark photon also couples to light dark matter particles with couplings typically of ${\cal O}(1)$, its total decay width will be a few orders of magnitude larger, therefore lowering the visibility of the resonant signal.

Quantitatively, if we require that we generate the same size of the dark photon signal cross sections when $g_{\chi}$ is turned on, the CMS constraints on $\epsilon$ are relaxed by one to two orders of magnitude~\cite{Felix:2025afw} and are then compatible with the determinations from electroweak precision observables~\cite{Curtin:2014cca, Loizos:2023xbj}, electron-nucleon deep-inelastic scattering~\cite{Kribs:2020vyk, Thomas:2021lub, Yan:2022npz}, as well as our best fit result. We would suggest that future direct experimental searches at $e^+ e^-$ and hadron colliders aim to constrain signals associated with a broad dark boson resonance.

%%%%%%%%%%%%%%%%%%%%%%%%%%%%%%%%%%%%%%%%%%%%%%%%%%%%%%
\section{Implications of a dark photon}
\label{sec:implications}

\subsection{Parity-violating electron scattering}

Parity-violating electron scattering (PVES) offers a powerful tool for testing the SM and probing new physics by precisely measuring the fundamental couplings $C_{1q}$, $C_{2q}$ and $C_{3q}$, which are defined as the products of weak couplings to the electron and quarks~\cite{Erler:2013xha}.

In the case of deep inelastic scattering (DIS), especially from a deuteron target, the PVES asymmetry and the lepton charge asymmetry provide direct connection to these couplings~\cite{Zheng:2021hcf}
\begin{eqnarray}
\label{eq:A_d}
A_d^{e^-_R - e^-_L} &=& \frac{3 G_F Q^2}{10 \sqrt{2} \pi \alpha_{em}} \Big[ (2 C_{1u} - C_{1d}) + R_V Y (2 C_{2u} - C_{2d}) \Big]\, ,\nonumber\\
A_d^{e^+ - e^-} &=& -  \frac{3 G_F Q^2 Y}{2 \sqrt{2} \pi \alpha_{em}} \frac{R_V (2 C_{3u} - C_{3d})}{5 + 4 R_C + R_S}\, .
\end{eqnarray}
As a consequence of the nonzero axial-vector couplings, the dark photon will also contribute to the PVES asymmetries. The differential cross section of longitudinally polarised electron scattering off unpolarised target is~\cite{Thomas:2022qhj}
{\small
\begin{eqnarray}
\frac{d^2 \sigma}{dx dy} 
&=& \frac{4\pi \alpha^2 s}{Q^4}
\Big(
[x y^2 F_1^{\gamma} + f_1(x,y) F_2^{\gamma}] \nonumber\\
&& - \frac{1}{\sin^2 2\theta_W}\frac{Q^2}{Q^2 + M_Z^2} (C_{Z,e}^v - \lambda C_{Z,e}^a)
 [x y^2 F_1^{\gamma Z} + f_1(x,y) F_2^{\gamma Z} - \lambda x y (1-\frac{y}{2}) F_3^{\gamma Z}]  \nonumber\\
&&
-  \frac{1}{\sin^2 2\theta_W} \frac{Q^2}{Q^2 + M_{A_D}^2} (C_{A_D,e}^v - \lambda C_{A_D,e}^a)
[x y^2 F_1^{\gamma A_D} + f_1(x,y) F_2^{\gamma A_D} - \lambda x y (1-\frac{y}{2}) F_3^{\gamma A_D}]
\Big) \, ,\nonumber\\
\end{eqnarray}}
where  $f_1(x,y) = 1 - y - xyM/2E$ and $\lambda = + 1 (-1)$ represents positive (negative) initial electron helicity.

By calculating relevant PVES asymmetries and the lepton charge asymmetry in Eq.~(\ref{eq:A_d}), we found that the total effect of the physical $Z$ and $A_D$ exchanges is given by the effective coupling~\cite{Thomas:2022qhj}
\begin{equation}
\label{eq:C1q}
C_{1q} = C^Z_{1q} + \frac{Q^2 + M_Z^2}{Q^2 + M_{A_D}^2} C^{A_D}_{1q} \equiv C^{\rm SM}_{1q} ( 1 + R_{1q} )\, ,
\end{equation}
where $R_{1q}$ characterises the dark photon correction to its SM prediction, which depends on the dark parameters $(\epsilon, M_{A_D})$ and the momentum transfer scale $Q^2$. $C_{2q}$ and $C_{3q}$ have similar forms to Eq.~(\ref{eq:C1q}).

\begin{figure}[!h]
\centering
\includegraphics[width=0.49\textwidth]{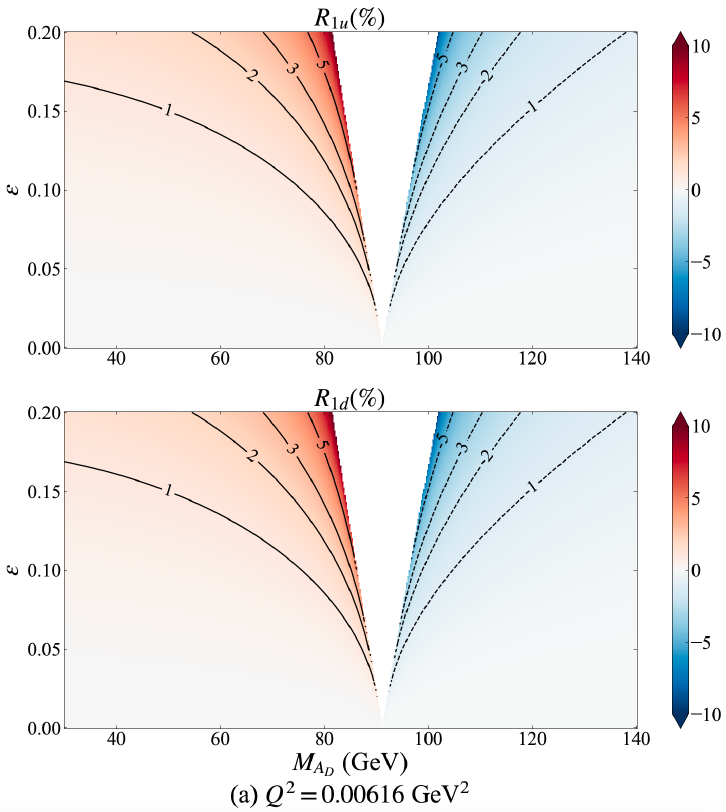}
\hspace*{0.1cm}
\includegraphics[width=0.49\textwidth]{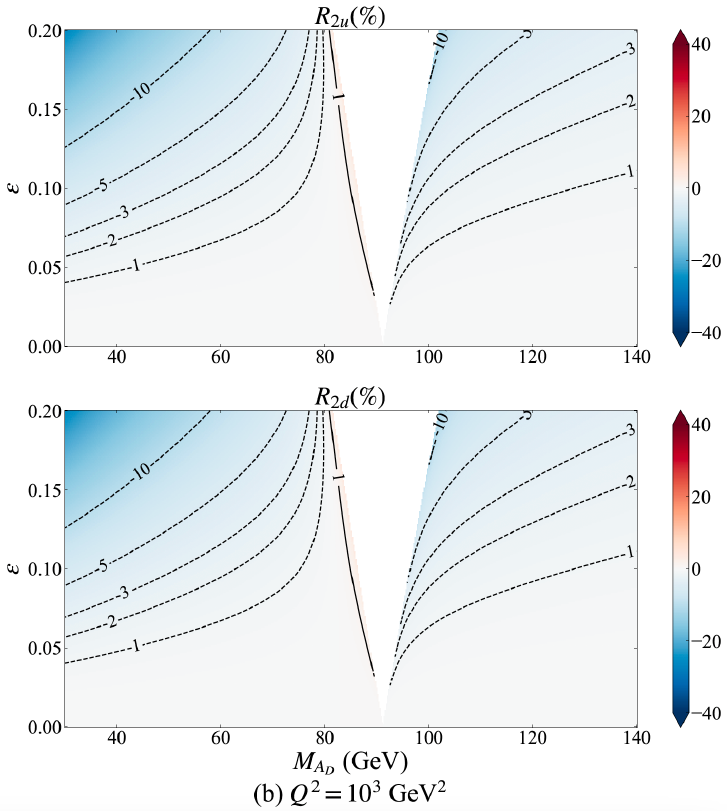}
\vspace*{-0.1cm}
\caption{ (a) The corrections to $C_{1q}$ at low scale $Q^2 = 0.00616\ {\rm GeV}^2$; (b) The corrections to $C_{2q}$ at high scale $Q^2 = 10^3\ {\rm GeV}^2$. The gap is the ``eigenmass repulsion" region in which the dark photon parameters are not accessible~\cite{Kribs:2020vyk}.}
\label{fig:R1-R2}
\end{figure}

The sensitivities of the correction factors to dark photon parameters are shown in Fig.~\ref{fig:R1-R2}~\cite{Thomas:2022qhj, Thomas:2025qbe}.
 At low scales, the corrections to $C_{1q}$ could be as large as $5\%$ when the dark photon parameters approach the ``eigenmass replusion" region. At a typical scale of $Q^2\sim 10\ {\rm GeV}^2$ relevant for the SoLID experiment~\cite{JeffersonLabSoLID:2022iod}, the sensitivities of $R_{3q}$ to the dark photon parameters are similar to those of $R_{1q}$ in Fig.~\ref{fig:R1-R2}. In DIS at very high scale, $Q^2 = 10^{3}\ {\rm GeV}^2$, of relevance to HERA or the EIC,  the dark photon could induce substantial corrections to $C_{2q}$, suggesting uncertainties as large as $10\%$ in the extraction of valence parton distribution functions.

\subsection{Rare kaon decay}

Among rare kaon decays, $K^+ \to \pi^+ \nu\bar{\nu}$ and $K_L \to \pi^0 \nu\bar{\nu}$ are the golden channels as their SM values of the branching fractions can be computed accurately. The SM predictions and the latest experimental results are summarised in Tab.~\ref{table:Braching}. Potential anomalies in these branching ratios would imply signals of new physics. 

\begin{table}[!h]
\begin{center}
\begin{tabular}{c|c|c}\hline\hline
 \   &   SM~\cite{Buras:2022wpw}     &   Experiment \\ \hline
 ${\rm Br}(K^+ \to \pi^+ \nu\bar{\nu})$  &  $(8.6\pm 0.42)\times 10^{-11}$   &  $(13.0^{+3.3}_{-3.0})\times 10^{-11}$~\cite{NA62:2024pjp}  \\ 
 ${\rm Br}(K_L \to \pi^0 \nu\bar{\nu})$  &  $(2.94 \pm 0.15) \times 10^{-11}$  &  $< 2.2 \times 10^{-9}$~\cite{KOTO:2024zbl} \\
 \hline 
\end{tabular}
\caption{The theoretical and experimental results of the branching fractions.}
\label{table:Braching}
\end{center}
\end{table}

It was expected that a light dark photon might lead to sizable modifications to the SM branching ratios because of the enhancement from its propagator, compared with that of the $Z$ boson. We~\cite{Wang:2023css} focused on the neutral channel~\cite{Buchalla:1995vs},
\begin{equation}
\label{eq:Br-KL}
    {\rm Br}(K_L \to \pi^0 \nu\bar{\nu}) = \kappa_L \left( \frac{{\rm Im} \lambda_t}{\lambda^5} X(x_t)\right)^2\, ,
\end{equation}
where the top quark contribution is dominant, with $\lambda_t = V^*_{t s} V_{t d}$ being the CKM factor, $\kappa_L = (2.231 \pm 0.013) \times 10^{-10} (\lambda/0.225)^8$ parameterizing the hadronic matrix element and here $\lambda$ is the 12 element of the CKM matrix. For the charged kaon decay one has to also consider the charm-quark contribution. 

When dark photon effects are included, the decay amplitude is~\cite{Wang:2023css}
\begin{equation}
\label{eq:X0}
X^{(0)}(x_q,y_l) 
= - \frac{2 m^2_W}{\cos^2\theta_W} \frac{ C_{Z, \nu_l}}{k^2 - M^2_Z} C^{(0)}_{Z}(x_q) 
- \frac{2 m^2_W}{\cos^2\theta_W} \frac{ C_{A_D, \nu_l}}{k^2 - M^2_{A_D}} C^{(0)}_{A_D}(x_q) 
- 4 B^{(0)}(x_q,y_l) \, ,
\end{equation}
where $C^{(0)}_{Z(A_D)}(x_q)$ and $B^{(0)}(x_q,y_l)$ are loop functions associated with the $Z(A_D)$-penguin and the $W$-box diagrams, respectively.  
The dark photon effect can be characterised by a correction factor to the SM branching ratio,
\begin{equation}
\frac{{\rm Br}(K_L \rightarrow \pi^0 \nu\bar{\nu})}{{\rm Br}(K_L \rightarrow \pi^0 \nu\bar{\nu})|_{\rm SM}} = 1 + R_L \, , 
\end{equation}
which is independent of $\kappa_L$, $\lambda_t$, and $\lambda$. 

\begin{figure}[!htpb]
\centering
\includegraphics[width=0.7\textwidth]{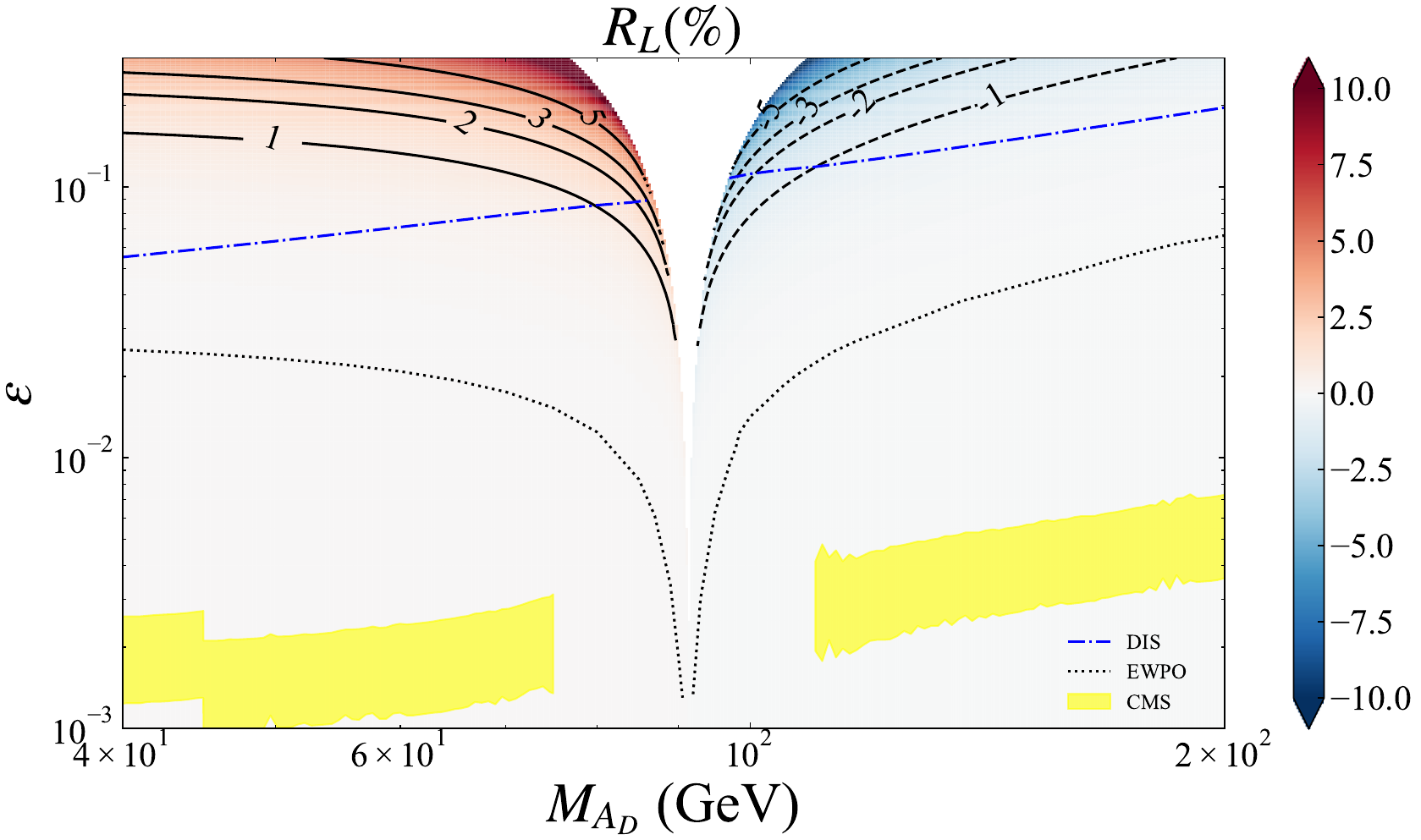}
\vspace*{-0.1cm}
\caption{ The dark photon corrections to the SM branching ratio ${\rm Br}(K_L \to \pi^0 \nu\bar{\nu})$.}
\label{fig:R_KL}
\end{figure}

The sensitivity of $R_L$ to the dark photon parameters is shown in Fig.~\ref{fig:R_KL}~\cite{Wang:2023css}. Contrary to the naive expectation, we found that in the allowed region of the dark photon parameter space---below the existing exclusion limits from electroweak precision observables~\cite{Curtin:2014cca}, the CMS collaboration~\cite{CMS:2019buh}, and even the weakest constraints from electron-nucleon deep-inelastic scattering~\cite{Thomas:2021lub}---the dark photon can only lead to less than $1\%$ corrections to the SM branching ratio, which are way to small compared with the experimental accuracy. The reason is that, the dark photon couplings to the neutrinos in Eq.~(\ref{eq:X0}) not only depend on the mixing parameter $\epsilon$ but also scale as $M_{A_D}/M_Z$, therefore eliminating the enhancement associated with its propagator.

If a sizable anomaly in the branching ratio were to be confirmed by experiments in the future, the dark photon model might still be a viable if the dark photon coupled to light dark matter particles, in addition to the neutrino final states.

\subsection{Electroweak precision observables}

The dark photon is also an appealing portal that could potentially connect dark and ordinary matter~\cite{Fabbrichesi:2020wbt, Filippi:2020kii}. We take Dirac fermion dark matter as an example, by adding an interaction term ${\cal L}_{\chi} = g_{\chi} \bar{\chi} \gamma^{\mu} \chi A'_{\mu}$ to Eq.~(\ref{eq:L_DP}). We found that the $Z$ boson will also couple to the dark matter particle $\chi$ as a result of $\bar{Z} - A'$ mixing, with the physical coupling being $C_{Z, \chi\bar{\chi}} = g_{\chi} \sin\alpha / \sqrt{1- \epsilon^2/\cos^2\theta_W}$. 
As a result, the total decay width of the $Z$ boson will receive an extra contribution, in addition to the SM final states, for $m_{\chi} < M_Z/2$~\cite{Loizos:2023xbj},
\begin{eqnarray}
\Gamma_Z = \Gamma_{\rm had} + \Gamma_e + \Gamma_\mu + \Gamma_\tau + 3 \Gamma_{\nu} + \Gamma_{\chi}\ ,
\end{eqnarray}
where
\begin{equation}
\label{eq:Gamma_chi}
\Gamma_{\chi} = \frac{m_Z C^2_{Z,\chi\bar{\chi}}}{12\pi} \left( 1 + \frac{2 m^2_{\chi}}{m^2_Z} \right) \sqrt{1 - \frac{4 m^2_{\chi}}{m^2_Z}}\, .
\end{equation}
This allows us to place constraints on $\epsilon$ and $g_{\chi}$ independently from an analysis of electroweak precision observables.

Following the procedure performed in Ref.~\cite{Curtin:2014cca}, we chose the convenient set of free parameters 
\begin{equation}
m_h,\ m_Z,\ m_t,\ \alpha_s,\ \Delta\alpha^{(5)}_{\rm had}\, ,
\end{equation}
and define
\begin{equation}
    \chi^2= V \cdot cov^{-1} \cdot V\, ,\ \ cov = \Sigma_{\rm exp} \cdot cor \cdot \Sigma_{\rm exp}\, ,
\end{equation}
where $V = {\rm theory} (m_h,m_Z,m_t,\alpha_s,\Delta\alpha^{(5)}_{\rm had}, m_{A_D}, \epsilon,  g_{\chi}) - {\rm exp}$ denotes the difference vector between the theoretical predictions and the experimental measurements. 

\begin{figure}[!htpb]
\centering
\includegraphics[width=0.9\textwidth]{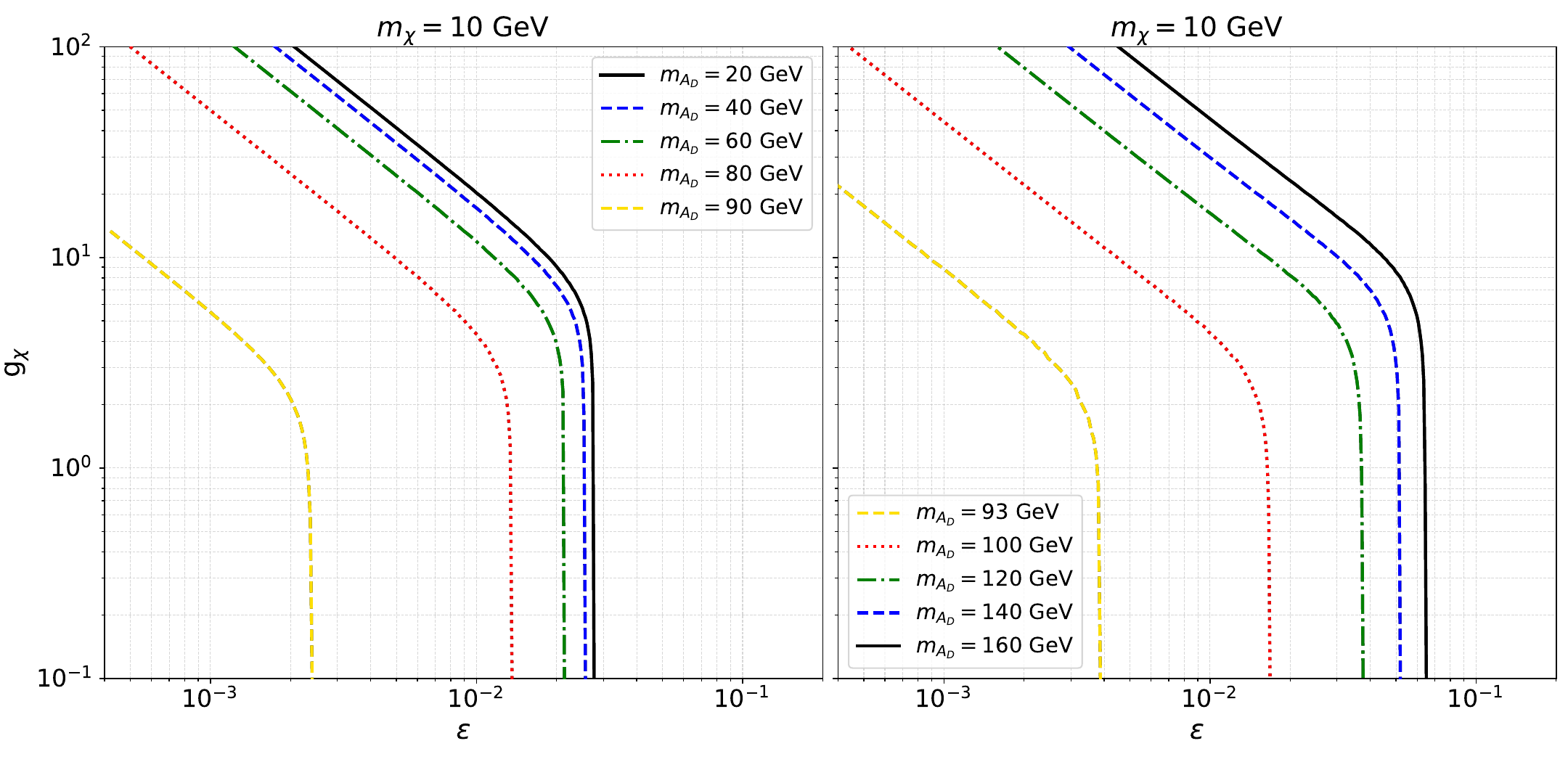}
\vspace*{-0.1cm}
\caption{ The 95\% CL exclusion constraints on dark parameters in the $g_{\chi} - \epsilon$ plane, using $m_W^{\rm PDG}$.}
\label{fig:g_chi_eps}
\end{figure}

We first performed the SM fit to the latest experimental data of 17 electroweak precision observables~\cite{ParticleDataGroup:2024cfk}, with the minimum $\chi^2$ being $\chi^2_{\rm SM}/d.o.f = 12.9/(17-5)$.
We then extended the analysis by including the dark photon and the dark fermion $\chi$. For a given dark photon mass, the minimum $\chi^2$ one can reach depends on $\epsilon$ and $g_{\chi}$. The exclusion limits on these two parameters at $95\%$ CL are set by
\begin{equation}
    \chi^2_{\rm DP}(\epsilon, g_{\chi}) - \chi^2_{\rm SM} = 5.99\, .
\end{equation}
The resulting constraints in the $g_{\chi} - \epsilon$ plane for $m_{\chi} = 10\ {\rm GeV}$ are shown in Fig.~\ref{fig:g_chi_eps}. The upper bounds are tightened as the dark photon mass approaches $M_Z$, and relaxed when $M_{A_D}$ goes above $M_Z$.
%

%%%%%%%%%%%%%%%%%%%%%%%%%%%%%%%%%%%%%%%%%%%%%%%%%%%%%
\section{Conclusion}
\label{sec:conclusion}
We performed a global fit to electron-nucleon deep inelastic scattering and related high-energy data without (baseline) and with the inclusion of a dark photon. We found a reduced $\chi^2$ in the latter case compared with the baseline result. The hypothesis test yielded a preference for the dark photon model as large as $6.5\ \sigma$. This result constitutes the first hint for the existence of a dark photon, albeit indirect.

We have explored the implications of a dark photon. In parity-violating electron scattering, the dark photon effects can be characterised by corrections to fundamental weak couplings in the SM, which could be as large as $5\%$ to $C_{1q}$ and $C_{3q}$ at low momentum scales and $10\%$ to $C_{2q}$ at high $Q^2$. The dark photon corrections to the Standard Model branching ratio of $K_L \to \pi^0 \nu\bar{\nu}$ are less than $1\%$ if we only take into account the neutrino final states, contrary to the naive expectation. We have also investigated the dark photon as a potential portal connecting to the dark matter sector, by introducing its coupling to dark Dirac fermion $\chi$. From an analysis of electroweak precision observables, we set upper limits directly on the coupling $g_{\chi}$. 

\acknowledgments{
We would like to thank Anthony Williams, Martin White, Wally Melnitchouk, Nobuo Sato, Nicholas Hunt-Smith, Bill Loizos, and Jake Felix for fruitful collaborations. This work was supported by The University of Adelaide and The Australian Research Council through the Centre of Excellence for Dark Matter Particle Physics (CE200100008).
}

\bibliographystyle{JHEP}
\bibliography{bibliography}

\end{document}